\documentstyle[aps,epsfig,prb,preprint]{revtex}

\newcommand{\re}{{\rm e}}

\newcommand{\erg}{{\rm erg}}
\newcommand{\cm}{{\rm cm}}
\newcommand{\nm}{{\rm nm}}

\newcommand{\Kcr}{K_{\rm cr}}
\newcommand{\M}{{\bf \it M}}
\newcommand{\Ms}{M_{\rm s}}
\newcommand{\Hv}{{\bf \it H}}

\begin{document}
%\draft

\title{Effect of crystalline disorder on magnetic switching in small magnetic cells}
\author{Daniel Braun\footnote{ Corresponding author. Fax: +1 914 945
4421; email: v2braun@us.ibm.com}}
\address{Infineon Technologies, 2070 State Route 52, Hopewell Junction, NY, 12533\\
MRAM Developement Alliance, IBM/Infineon Technologies, IBM Semiconductor
Research and Developement Center, 2070 State Route 52, Hopewell Junction,
NY, 12533}

\maketitle
\widetext
\centerline{\today}
\begin{abstract}
\begin{center}
\parbox{14cm}{I present a study of the
influence of disorder  in
thin magnetic films on the switching behavior of small magnetic cells of well
defined shape and size. The disorder considered arises from  randomly
oriented crystalline grains of different shape, size, and crystalline 
orientation which gives rise to locally
fluctuating intrinsic anisotropy directions and strengths. The study
comprises a theoretical investigation of a 
disordered Stoner Wohlfarth model, as well as  micromagnetic
simulations.  I show that the
fluctuations in the total anisotropy and therefore in the switching fields
are controlled by a single dimensionless parameter. The theoretical findings
are well confirmed by micromagnetic simulations of many different
samples.
}
\end{center}
\end{abstract}
%}

%\pacs{72.15.Rn, 73.23.-b, 72.80.Ng}
%\begin{multicols}{2}
\section{Introduction}
With the development of integrated magneto-resistive memory devices the need
to produce 
very many magnetic cells with reproducible magnetic switching behavior has
arisen. As the dimensions of these devices reach orders of a few 100nm,
the switching behavior is to good approximation that of a single domain
particle, and mainly controlled by the total magnetic
anisotropy. The magnetic  anisotropy contains contributions from various
sources, but for such small devices the contributions from the
shape anisotropy and from the crystalline (i.e.~material) anisotropy
dominate. For a	 
given magnetic material, the
former can be controlled to a good degree by the aspect ratio and the
thickness of the magnetic cells, while the latter depends on the
actual microscopic structure. Typically magnetic films are
polycrystalline with grain sizes of the order of a few nm to a few
10nm. The grains are oriented randomly. Since the preferred directions
of the crystalline anisotropy are defined by the crystal axes,
this leads to locally varying anisotropy contributions. Depending on the
relative strength of shape anisotropy and crystalline anisotropy, one may
therefore 
expect a more or less pronounced random component in the switching field. It
is the purpose of this paper to assess quantitatively how much 
fluctuation one may expect depending on the various parameters in the
problem. 

A lot of work in material science has been done over many years to examine
how magnetic parameters are influenced by the composition and micro
structure of thin films
\cite{Platt00,Shan99,Weber96,McKinlay96,Johnson95,Kim99,Varga98,Zhou00,Jang97}.Models
of disordered ferromagnets were 
studied analytically by Ignatchenko and co--workers in the late 1970s and
early 1980s (see \cite{Ignatschenko82} and references therein) as well as by
Kronm\"uller and co--workers \cite{Kronmueller77}. They   derived
``laws of approach of the magnetization to saturation''. Such laws were used
as early as 1931 for the characterization of materials \cite{Akulov31}. The
laws derived by Ignatchenko et al.~allowed to
extract quantitative information about the correlations of local magnetic
anisotropy \cite{Ignatschenko78}.   The focus of these
works was, however, on average 
macroscopic characteristics, like coercive fields or remanent
magnetizations, and not on fluctuation of switching fields in small magnetic
elements. Ref.~\cite{Zheng99} deals with the latter problem by micromagnetic
simulations, but without a simple model presented the physical insight and
predictive 
power of this analysis is limited. Also in the context of recording media,
granular materials were extensively simulated \cite{Hurley99}. 
Experimentally it is very difficult to
distinguish the contributions of different  origins to fluctuations in
switching fields. In particular, fluctuations in the shapes of magnetic
cells may easily mask fluctuations arising from the material. 

In the following I will introduce a disordered Stoner Wohlfarth model,
a simple model that allows to address the question of disorder induced
fluctuations of the
switching fields analytically. I will show that within this model the joint
probability distribution $P(K,\gamma)$ for the total anisotropy strength $K$
and the 
overall preferred direction $\gamma$ is controlled by a single dimensionless
parameter $\nu$. If we denote by $K_1$ the uniaxial shape
anisotropy, by $K_{\rm cr}$ the crystalline (i.e. bulk single crystal) anisotropy, by $N$ the typical number of crystallites in the sample, and
by $\alpha$ a number of order unity that depends on the shape of the
crystallites, the parameter $\nu$ is given by
\begin{equation} \label{nu}
\nu=\frac{N}{\alpha}\left(\frac{K_1}{K_{\rm cr}}\right)^2\,.
\end{equation}
The fluctuations of the switching field decay like
$1/\sqrt{\nu}$ for large $\nu$. All of this will be derived in 
Sec.\ref{chap.ana}. Sec.\ref{chap.num} is devoted to numerical
verification by micromagnetic modeling, and Sec.\ref{chap.concl} contains
the conclusions.

\section{The disordered Stoner Wohlfarth Model}\label{chap.ana}
Suppose the magnetic cell consists of $N$ crystallites, with respective volumes
$V_l$, easy axis direction $\gamma_l$ ($l=1,\ldots,N$) and crystalline
anisotropy $K_{\rm cr}$. For the moment let us consider the simplest case in
which 
each crystallite gives rise to a uniaxial anisotropy. Further assume that
the magnetization $\M$ is uniform 
across the sample, with components uniquely specified by the angle $\theta$,
$M_x=M_s \cos\theta$, $M_y=M_s \sin\theta$, where $\M_s$ is the saturation
magnetization. This approximation works well for sub-$\mu$m sized magnetic 
cells, which are too small to hold domain walls. In an external magnetic field in the cell plane,
$\Hv=(H_x,H_y)$, the total energy density then reads
\begin{equation} \label{e}
E/V=K_1\sin^2\theta+K_{\rm cr}\sum_{l=1}^Nv_l\sin^2(\theta-\gamma_l)-H_x\Ms\cos\theta-H_y\Ms\sin\theta\,, 
\end{equation}
where $v_l=V_l/V$ are the volume fractions of the crystallites ($V$ denotes
the total volume of the magnetic cell). Note that the
forms of the crystallites do not enter at this point. The first and the
last two terms on the right hand side 
correspond to the ordinary Stoner Wohlfarth model \cite{Stoner48}.

In order to figure out the total anisotropy resulting from eq.(\ref{e}) we
need to understand how the 
different  anisotropy terms add up. Let us start by looking at just two
uniaxial anisotropies $K_1$ and $K_2$ with 
preferred angles  $\gamma_1=0$ and $\gamma_2$. The energy density
is then given by  
\begin{equation} \label{e2}
E(\theta)/V=K_1\sin^2\theta+K_2\sin^2(\theta-\gamma_2)\,.
\end{equation}
We can rewrite this as 
\begin{equation} \label{e3}
E(\theta)/V=\epsilon+K\sin^2(\theta-\gamma)\,,
\end{equation}
where $\epsilon$ is a constant independent of $\theta$, and $K$ and $\gamma$
are the new anisotropy strength and preferred direction, respectively. By
expanding the $\sin^2$ terms in both eqs.(\ref{e2}) and (\ref{e3}) one
easily convinces oneself that the
three parameters $\epsilon$, $K$ and $\gamma$ are related to $K_1$, $K_2$
and $\gamma_2$ by the three equations
\begin{eqnarray}
K_1+K_2\cos 2\gamma_2&=&K\cos 2\gamma\,,\\
K_2\sin 2\gamma_2&=&K\sin 2\gamma\,,\\
K_2\sin^2\gamma_2&=& K\sin^2\gamma+\epsilon\,.
\end{eqnarray}
I will use the convention that all anisotropy
constants are positive ($0\le K_1,K_2,K\le \infty$) and all uniaxial
preference angles are counted in the interval $-\pi/2\le \gamma_2,\gamma<
\pi/2$. The system of  equations is then solved uniquely for all $K_1,K_2$,
and $\gamma_2$ by 
\begin{eqnarray}
\gamma&=&\frac{1}{2}{\rm sign}(\gamma_2)\arccos\left(\frac{K_1+K_2\cos
2\gamma_2}{K}\right)\,,\label{gamma}\\
K&=&\sqrt{K_1^2+K_2^2+2K_1K_2\cos 2\gamma_2}\,,\label{K}
\end{eqnarray}
and a third equation determining $\epsilon$, which is, however, irrelevant
for the following. Note that the $\arccos$ function returns a value in the
interval $0\ldots\pi$, which by the prefactor $1/2$ is remapped to the
interval $0\ldots \pi/2$.  Since
$\gamma_1$ was chosen as zero, the second angle 
determines the sign of the angle of the overall preferred
direction. Fig.\ref{fig.gamma} shows how the overall angle depends on
$\gamma_2$ for various values of the ratio $r=K_2/K_1$. 
\noindent
\begin{minipage}{7.0truein}
\begin{center}
\begin{figure}[h]
\epsfig{file=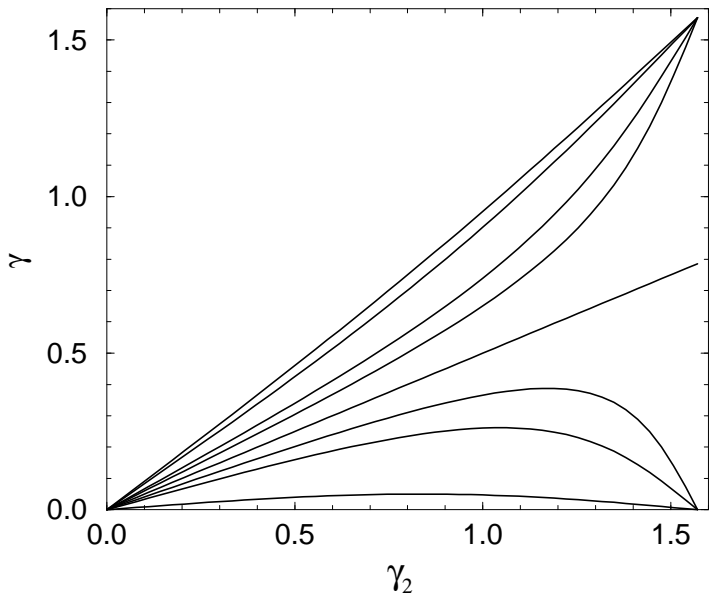,width=8cm,angle=270} \\[0.2cm]
\caption{Resulting angle from adding two uniaxial anisotropies
(eq.\ref{gamma}). For $r=K_2/K_1<1$, the maximum angle reached is smaller
than $\pi/4$. Only if $K_2$ dominates ($r>1$), the angle $\pi/2$ can be
reached. The values of $r$ chosen are 0.1 (the curve with the smallest
slope at $\gamma_2=0$), 0.5, 0.7, 1.0 (a straight line with slope 1/2), 1.5,
2.0, 5.0, 
and 10.0 (almost a straight line with slope 1.0). Only the first quadrant
is shown, as $\gamma$ is an odd function of $\gamma_2$.}\label{fig.gamma}  
\end{figure}
\end{center}
\end{minipage}
\vspace*{0.1cm}

For $K_2/K_1<1$ the
maximum angle reachable for the new overall preferred direction is smaller
than $\pi/4$ --- the $K_1$ term always dominates and keeps the preferred
direction close to zero. For $K_1=K_2$, an overall angle $\pi/4$ can be
reached at $\gamma_2=\pi/2$, but eq.(\ref{K}) shows that then the 
overall strength $K$ goes to zero: With two orthogonal preferred axes with
anisotropy of equal strength, the total anisotropy vanishes indeed. For
$K_2>K_1$, the overall preferred angle is dominated by the second anisotropy
and therefore angles beyond $\pi/4$ can be reached.

Concerning the total strength of the anisotropy, eq.(\ref{K}) tells us that
$K$ is obtained similarly as if two vectors with lengths and directions
$K_1,K_2$ and directions $0,\gamma_2$ were added --- with the only
difference that the angle between the two vectors has to be replaced by
{\em twice} its value in the rule for standard vector addition. Thus, the
non--linear addition of uniaxial anisotropy terms is, for what concerns the
resulting anisotropy strength, replaced by a linear vector addition, where
the length of the vectors is given by the strengths of the anisotropy, and
their enclosed angle is twice the angle between the original preferred
directions. 

Coming back to the disordered Stoner--Wohlfarth model, eq.(\ref{e}), we now
understand that the sum of anisotropy energies with random strength and
preferred directions, i.e. the intrinsic anisotropy
$K_i=\Kcr\sum_{l=1}^Nv_l\sin^2(\theta-\gamma_l) $, 
corresponds to 
a random walk in the plane: For the vector addition the relative angles are
uniformly distributed in the interval $[-\pi,\pi[$, and the length of  
step $l$ is the random term $\Kcr v_l$. The unusual addition (\ref{gamma}) of
angles not 
withstanding, the resulting final orientation of the vector in the
random walk is distributed uniformly
over $[-\pi,\pi[$, for all 
$K_i$  since
each single angle is. The random walk leads to a distribution of $K_i$ that
is for $N\gg 1$ well approximated by a  Gaussian centered at $K_i=0$ and
with a variance
$\sigma^2=\langle K_i^2\rangle=N \Kcr^2\langle v_i^2\rangle$. If we assume
that the 
typical volume fraction is given by a typical crystallite dimension $a_l$ as
$\langle v_l^2\rangle^{1/2}=\sqrt{\alpha}\langle a_l^2\rangle/L^2=\sqrt{\alpha}/N$, where $\alpha$
is a numerical 
prefactor depending on the distribution of crystallite areas (and thus on
the shape of the crystallites) we obtain the scaling behavior 
$\sigma=\sqrt{\alpha}\Kcr/\sqrt{N}$, and the joint probability
distribution $P(K_i,\gamma_i)$  of
the intrinsic anisotropy and the preferred direction,
\begin{equation} \label{PKi}
P_i(K_i,\gamma_i)=\frac{\sqrt{2}}{\sqrt{\pi}\sigma}\exp\left(-\frac{K_i^2}{2\sigma^2}\right)\frac{1}{\pi} 
\end{equation}
for $-\pi/2\le \gamma_i<\pi/2$ and $0\le K_i\le\infty$.

In order to determine the distribution of the total anisotropy, it is most
convenient to start with an expression for the joint distribution of
$K^2$ and the total preferred angle $\gamma$, employing once more the law
for adding uniaxial anisotropies, eqs.(\ref{gamma}) and (\ref{K}). We have
\begin{eqnarray} \label{PK2gam}
P(K^2,\gamma)&=&\sqrt{\frac{2}{\pi}}\int_0^\infty\frac{dK_i}{\sigma}\int_{-\pi/2}^{\pi/2}\frac{d\gamma_i}{\pi}\delta\left(K_i^2+2K_1K_i\cos
2\gamma_i+K_1^2-K^2\right)\nonumber\\
&&\times\delta\left(\gamma-\frac{1}{2}{\rm sign}(\gamma_i)\arccos\left(\frac{K_1+K_i\cos2\gamma_i}{K}\right)\right)\exp\left(-K_i^2/(2\sigma^2)\right)\,.
\end{eqnarray}
The two integrals are easily performed. We can then transform the
distribution back to $P(K,\gamma)$, express $K$ with a dimensionless
parameter $k$ as $K=kK_1$, and thus arrive at the final form 
\begin{equation} \label{PKgamma}
P(K,\gamma)=\frac{1}{K_1}\sqrt{\frac{2\nu}{\pi^3}}k\frac{\re^{-(k^2+1-2k\cos2\gamma)\nu/2}}{\sqrt{k^2+1-2k\cos
2\gamma}} 
\end{equation} 
with the dimensionless parameter $\nu$ defined pre\-vious\-ly in eq.(\ref{nu}).
As is now obvious, the total distribution of anisotropies is uniquely
specified by this parameter $\nu$, and so are all statistical properties of
the switching fields. 

The distribution is 
centered around $k=1$ and $\gamma=0$, and in fact diverges for these values
for all parameters $\nu$. Typical values  of $\nu$ for magnetic cells may
vary over a large range. A rectangular 800x400x5 ${\rm nm}^3$ Permalloy cell
leads to a 
shape anisotropy $K_1=44.4\cdot 10^3{\rm erg/cm^3}$. Assuming a crystalline anisotropy $\Kcr=1000$
erg/$\rm{cm}^3$ and a typical crystallite size of 20nm, we
have $\nu=1.58\cdot 10^6$. Larger crystallites (say $\langle
a_i^2\rangle^{1/2}=50{\rm nm}$), 
a smaller aspect ratio, and thinner films may reduce this value. A
rectangular 500x400x3 ${\rm nm}^3$ cell 
has $\nu=10.2\cdot 10^3$.

The distribution $P(K,\gamma)$ is most relevant for rotational remanence
experiments on arrays of nominally identical magnetic cells \cite{rotrem}. In
these 
experiments a strong magnetic field is applied at an angle $\beta$ relative
to the nominal easy axis of the cells (as defined by shape anisotropy). Then
the magnetic field is switched off, with the direction of the field
preserved until zero field is reached, and one measures the remanent
magnetization 
along the nominal easy axis as function 
of the angle
$\beta$. In the strong magnetic fields (i.e.~field values outside all
astroids of the cells), a cell will always align to good approximation to
the field. But when the field is switched off, cells that saw a positive
field component
along their actual easy axis will remain in a state magnetized along their
positive actual easy axis, while those which saw a negative field component
relative to their actual easy axis will fall into a state magnetized along
their negative actual easy axis. Thus, all that matters
is the distribution of preferred angles $\gamma$, but not the strength of
the anisotropy.  Let us  call ``$x$--axis''the nominal 
easy direction, $M_c$ the average total magnetic moment of a
single 
cell along the $x$--axis at zero field, and $n$ the number of cells in the
array.  If all
cells had  $\gamma=0$
there would be a sharp jump of the total  $M_x$ of the array at
$\beta=\pi/2$ from $M_x=nM_c$ to 
$M_x=-nM_c$, if $\beta$ is cranked up from zero to $\pi$. A finite
width of the distribution of $\gamma$ is reflected directly in the
width of the transition. Integrating out $K$ in (\ref{PKgamma}) we find the
distribution of the preferred angles alone. Fig.\ref{fig.Pofgamma} shows the
result obtained from numerically integrating
\begin{equation} \label{Pgamma}
P(\gamma)=\int_0^\infty dK P(K,\gamma)\,.
\end{equation}
Note the rather non--Gaussian profiles, in particular the pronounced cusps
at zero angle,
even for very small values of $\nu$ where the distribution is almost
homogeneous over the entire angle interval $-\pi/2\ldots \pi/2$.  

\noindent
\begin{minipage}{7.0truein}
\begin{center}
\begin{figure}[h]
\epsfig{file=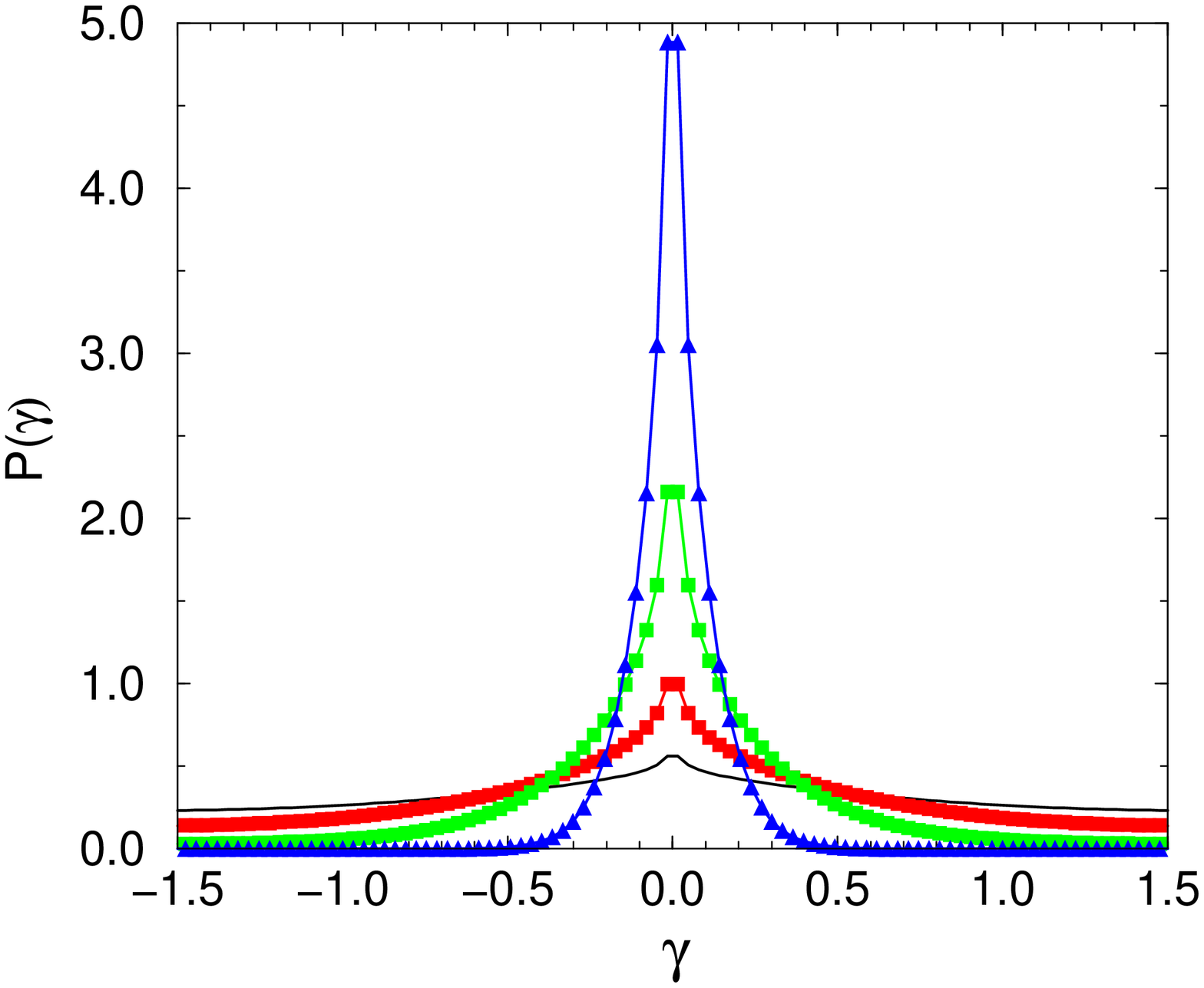,width=7cm,angle=270} 
\epsfig{file=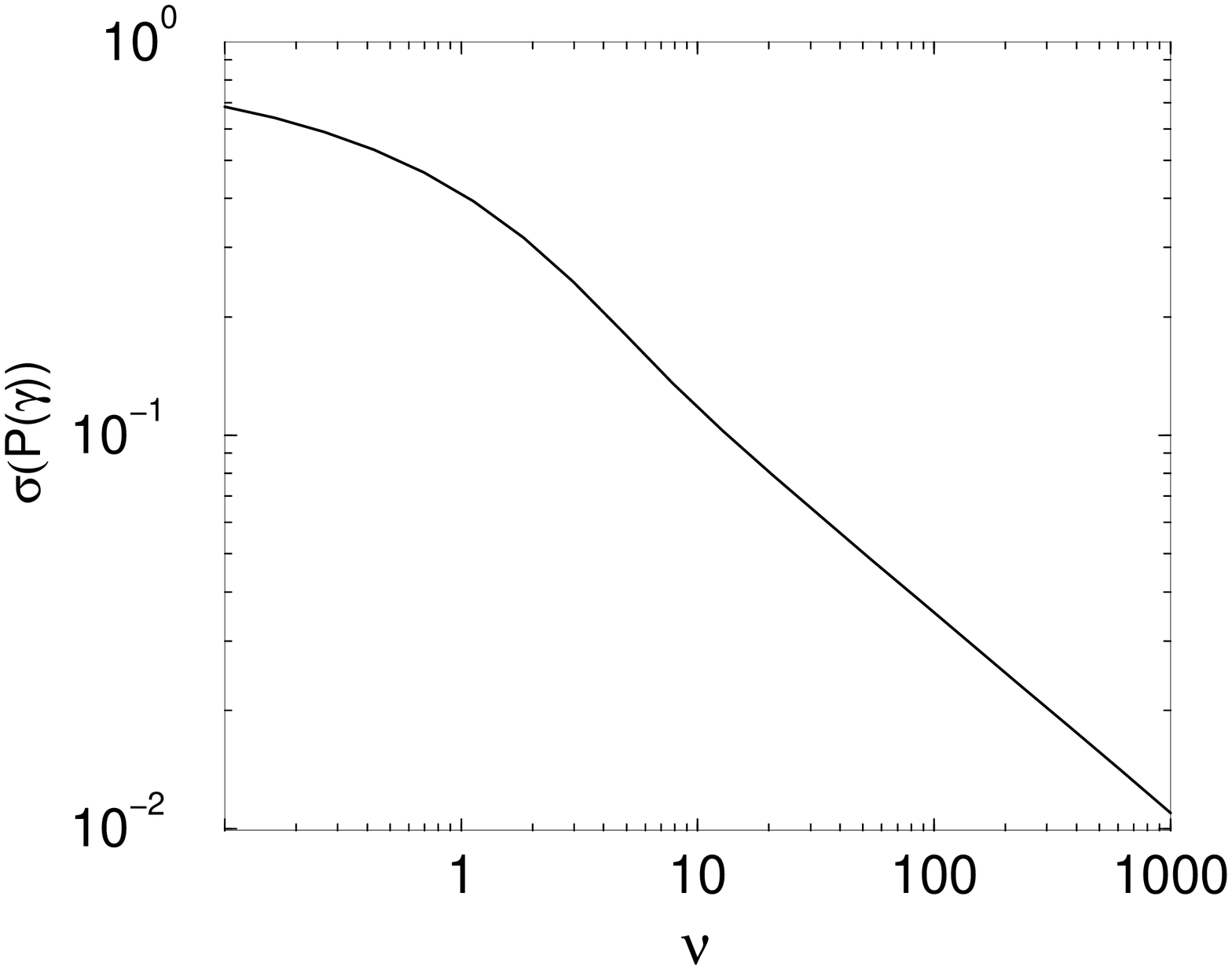,width=7cm,angle=270} \\[0.2cm]
\caption{{\em Left}: Distribution $P(\gamma)$ of the preferred angle, after
integrating 
out the anisotropy strength $K$. The parameter
$\nu$ ranges from $\nu=0.01$ (no symbols, flat curve) to $\nu=10$
(triangles). {\em Right}: Scaling of the width of the distribution $P(\gamma)$ with the
parameter $\nu$.}\label{fig.Pofgamma}  
\end{figure}
\end{center}
\end{minipage}
\vspace*{0.1cm}

The standard deviation of the distribution scales
as $1/\sqrt{\nu}$, as is shown in Fig.\ref{fig.Pofgamma}.
For comparison with actual experiments it is useful to note, however, that
the cells in an array are typically {\em not} identical, even concerning
their geometry. Fluctuations in the shape due to lithography errors and etch
processes will lead to additional fluctuations in $K_1$ that can mask
fluctuations due to the intrinsic material anisotropy.

\subsection{Switching field fluctuations}
Let us now have a look at the consequences for the distribution of switching
fields.  As is well known, the energy density (\ref{e}) without the
fluctuating term ($\Kcr=0$) leads to a stability region in the $(H_x,H_y)$
plane given by the astroid \cite{Stoner48}
\begin{eqnarray} 
h_{x,0}(\theta)&\equiv& \frac{M_s H_{x}(\theta)}{2K_1}=-\cos^3\theta\,,\label{ast0x}\\
h_{y,0}(\theta)&\equiv& \frac{M_s
H_{y}(\theta)}{2K_1}=\sin^3\theta\,.\label{ast0y} 
\end{eqnarray} 
The two equations are obtained by setting simultaneously $\partial_\theta
E(\theta)/V=0$ and $\partial_\theta^2
E(\theta)/V=0$, which leads to the field combinations
$(H_x(\theta),H_y(\theta))$ where the nature of 
equilibrium points changes from stable to unstable (or from unstable to
stable).   
A magnetic cell in a stable state with $M_x<0$ switches to a state $M_x>0$
if a field combination is 
applied that lies outside of the boundary (\ref{ast0x}), (\ref{ast0y}) in the
positive half plane $h_x>0$ (and correspondingly for $M_x<0$).

Including the random crystalline contributions in eq.(\ref{e}) we now have
\begin{equation} \label{etot}
E(\theta)/V=K\sin^2(\theta-\gamma)-M_s H_x\cos\theta-M_s H_y\sin\theta
\end{equation}
for the  total energy density
for a given cell (up to the irrelevant global shift $\epsilon$), where $K$
and $\gamma$ are distributed according to (\ref{PKgamma}). Since the easy
axis direction defines the orientation of the coordinate frame relative to
which the Stoner--Wohlfarth astroid is measured, it is clear even without
calculation that the energy density (\ref{etot}) leads for each $K$ and
$\gamma$ again to an ideal astroid that is rotated by the angle
$\gamma$ and changed in size by a factor $K/K_1$ (if magnetic fields are
still measured in units of $2K_1/M_s$, the normalization used in
eqs.(\ref{ast0x},\ref{ast0y})). In the explicit form of the stability
boundary derived again from $\partial_\theta E(\theta)=0=\partial_\theta^2 E(\theta)$,
\begin{eqnarray}
h_{x}(\theta)&=& \frac{M_s H_{x}(\theta)}{2K_1}=-\frac{K}{4K_1}\left(\cos(3\theta-2\gamma)+3\cos(\theta-2\gamma)\right)\,,\label{astx}\\
h_{y}(\theta)&=& \frac{M_sH_{y}(\theta)}{2K_1}=\frac{K}{4K_1}\left(-\sin(3\theta-2\gamma)+3\sin(\theta-2\gamma)\right)\,,\label{asty} 
\end{eqnarray}
the rotation of the astroid is somewhat obscured by the fact that $\theta$
is not the polar angle of the astroid, but just a parameter in a parametric
representation. One easily convinces oneself, though, that acting on
$(h_{x,0},h_{y,0})$ with a rotation matrix corresponding to an angle
$\gamma$ and with an overall factor $K/K_1$ reproduces
(\ref{astx}),(\ref{asty}). 

Eqs.(\ref{astx}) and (\ref{asty}) can be used together with
(\ref{PKgamma}) to calculate an average astroid as well as fluctuations
around it. It turns out, however, that for this purpose  it is more
convenient to go back one 
step in the calculation and keep the total intrinsic anisotropy $K_i$
separate from the deterministic anisotropy $K_1$, since $K_i$ is
simply Gaussian distributed, eq.(\ref{PKi}). This will prove useful in
Sec.\ref{sec.cubic} as well, where we will look at the combination of random
cubic crystalline anisotropy with uniaxial shape anisotropy. The latter
situation will not even allow for defining an overall anisotropy with a
single anisotropy constant. Let us therefore start from an energy density
\begin{equation} \label{eKi}
E(\theta)/V=K_1\sin^2\theta+K_i\sin^2(\theta-\gamma_i)-M_sH_x\cos\theta-M_sH_y\sin\theta
\end{equation}
with $K_i,\gamma_i$ distributed according to (\ref{PKi}). If we express
$K_i$ in units of $K_1$, $K_i=k_iK_1$, we arrive at
\begin{eqnarray} 
h_{x}(\theta)&=& \frac{M_s H_{x}(\theta)}{2K_1}=-\frac{k_i}{4}\left(\cos(3\theta-2\gamma_i)+3\cos(\theta-2\gamma_i)\right)-\cos^3\theta\,,\label{astix}\\
h_{y}(\theta)&=& \frac{M_s
H_{y}(\theta)}{2K_1}=-\frac{k_i}{4}\left(\sin(3\theta-2\gamma_i)-3\sin(\theta-2\gamma_i)\right)+\sin^3\theta\,.\label{astiy}  
\end{eqnarray} 
Before presenting the results for the mean values and fluctuations of the
switching fields, let me note that depending on the measurement performed
different ways of averaging might be relevant. If we are
interested in the fluctuation of the 
switching field for a given {\em direction} of the applied field  and if
the fluctuations are small, i.e.~$\nu\gg 1$, the fluctuations for a  
fixed parameter $\theta$ are relevant. The parameter $\theta$ will
be given to good approximation for all astroids by the unperturbed astroid
(\ref{ast0x}), (\ref{ast0y}),  
\begin{equation} \label{thetaav}
\theta=\arctan\left(\frac{H_y}{H_x}\right)^{1/3}\,.
\end{equation}
However, when the fluctuations are larger, or if we are interested in swept
astroids where $H_y$ is kept fixed and $H_x$ is swept, so that by definition
all 
fluctuations are  in $H_x$, the situation is much more
complex. One will then have  to calculate the relevant $\theta$ for each
realization of 
the disorder separately. This will be discussed elsewhere \cite{Braun01.2}. 

Here I will  assume that the disorder induced fluctuations are 
small and that the point on the astroid where we are interested in the
variation of the switching fields is sufficiently well described by $\theta$
as obtained from eq.(\ref{thetaav}).  
Since $k_i$ enters only
linearly in (\ref{astix}), (\ref{astiy}) and is distributed Gaussian, and
since 
$\gamma_i$ is distributed uniformly over the interval $-\pi/2\ldots\pi/2$,
we find immediately  that the average astroid is the ideal astroid
given in eqs.(\ref{ast0x}),(\ref{ast0y}). The standard deviation of the
switching field in units of $H_c=2K_1/M_s$ is given by 
\begin{equation} \label{sigmah}
\sigma_{Hx,Hy}=\frac{H_c}{4}\frac{1}{\sqrt{\nu}}\sqrt{5\pm 3\cos 2\theta}\mbox{, }
\end{equation}
where $\sigma_{Hx}$ comes with the positive sign under the square root,
$\sigma_{Hy}$ with the negative.
Thus, the fluctuations in the switching field scale like
$1/\sqrt{\nu}$. For soft magnetic materials like permalloy and magnetic
cells with aspect ratios not too close to unity, $K_1$ is dominated by shape
anisotropy, which in turn is proportional to the aspect ratio over a wide
range. We conclude that the ``array quality factor''  \cite{Abraham00}
$H_c/\sigma_{Hx}$  for fixed cell width
should be proportional to the aspect ratio of the cells to the power $3/2$.

\subsection{Cubic Crystalline Anisotropy}\label{sec.cubic}
A material with cubic anisotropy has an energy density that depends
on the direction cosines $\alpha_x$, $\alpha_y$ and $\alpha_z$ of the
magnetization with the crystal axes according to
\begin{equation} 
E/V=C(\alpha_x^2\alpha_y^2+\alpha_x^2\alpha_z^2+\alpha_y^2\alpha_z^2)\,,
\end{equation}
where $C$ is the lowest order cubic anisotropy constant \cite{Hubert98}. I
will assume in the following that the crystallites all have a $z$--axis
perpendicular to the film, i.e. the films are supposed to be well ordered in
$z$--direction. This is a natural assumption for flat crystallites with lateral dimensions of a few 10nm
and films only a few nm thick, even though uniform distributions of the
crystal axis on
cones have been observed in 50nm thick films \cite{Jang97}. Furthermore, if
the  magnetization is effectively restricted to 
the plane of  
the film (for sufficiently thin films this is always the case), we have
$\alpha_z=0$. I parameterize the magnetization in the 
plane again by an angle $\theta$ with respect to the $x$--axis. We then have
$\alpha_x=\cos\theta$,  $\alpha_y=\sin\theta$, and the expression reduces to   
\begin{equation} \label{Ec}
E/V=\frac{C}{4}\sin^2 2\theta\,.
\end{equation}
Cubic anisotropy projected to a (001) crystal plane thus leads to a
four--fold (or bi--axial) symmetry in that plane. And instead of the four--fold
jagged astroid for purely uniaxial materials, the stability curve is now
eight--fold jagged. Note that the shape anisotropy of the cells is still
uniaxial, though. The disordered Stoner Wohlfarth model for materials with the
projected cubic crystalline anisotropy thus reads
\begin{equation} \label{ec}
E/V=K_1\sin^2\theta+\frac{C}{4}\sum_{i=1}^Nv_i\sin^2(2(\theta-\gamma_i))-H_x\Ms\cos\theta-H_y\Ms\sin\theta\,,
\end{equation}
where I assume again that  the orientations $\gamma_i$ of the crystallites
are distributed uniformly over the entire relevant interval,
i.e.~$-\pi/4\le\gamma_i<\pi/4$. 
For adding only cubic anisotropies the same rules (\ref{gamma}), (\ref{K})
apply 
as were derived from adding uniaxial anisotropies. Indeed, in the derivation
we can just replace $\theta\rightarrow 2\theta$, and replace $K_1$, $K_2$ by
two corresponding cubic anisotropy constants $C_1$ and $C_2$, and everything
else goes through as before. The same is true for the random walk
picture. Thus, many cubic anisotropy terms added up lead again to a cubic
anisotropy with a distribution 
of the overall $C_i$ and $\gamma$ given by $2P_i(C_i,\gamma)$, see
eq.~(\ref{PKi}). The prefactor two is due to the fact that $\gamma$ now
covers only half the previous interval. Things are different, however, when
we combine the total cubic anisotropy with the uniaxial shape
anisotropy. Obviously, the result will neither be a pure cubic anisotropy
nor a pure 
uniaxial anisotropy, but rather a sum of two such terms. 
The boundary of the stability region derived from  $\partial_\theta
E(\theta)=0=\partial_\theta^2 E(\theta)$, with 
\begin{equation} \label{eCi}
E(\theta)/V=K_1\sin^2\theta+\frac{C_i}{4}\sin^2(2(\theta-\gamma_i))-M_sH_x\cos\theta-M_sH_y\sin\theta 
\end{equation}
now reads
\begin{eqnarray}
h_{x}(\theta)&=& -\frac{M_s H_{x}(\theta)}{2K_1}=\frac{c_i}{8}\left(5\cos(3\theta-4\gamma_i)+3\cos(5\theta-4\gamma_i)\right)-\cos^3\theta\,,\label{astcix}\\
h_{y}(\theta)&=& \frac{M_s
H_{y}(\theta)}{2K_1}=\frac{c_i}{8}\left(5\sin(3\theta-4\gamma_i)-3\sin(5\theta-4\gamma_i)\right)+\sin^3\theta\,.\label{astciy}
\end{eqnarray}
Depending on the relative strength $c_i=C_i/K_1$ and orientation of the
total cubic intrinsic 
anisotropy, this boundary may be rather different from the ideal
Stoner--Wohlfarth astroid, as for example depicted
in Fig.\ref{fig.bound}. Depending on the parameters, little twists arise
that might not always be resolvable in experiments, and give the 
impression of astroids broadened in one direction, or of kinks in the
astroid sides.  
\noindent
\begin{minipage}{7.0truein}
\begin{center}
\begin{figure}[h]
\epsfig{file=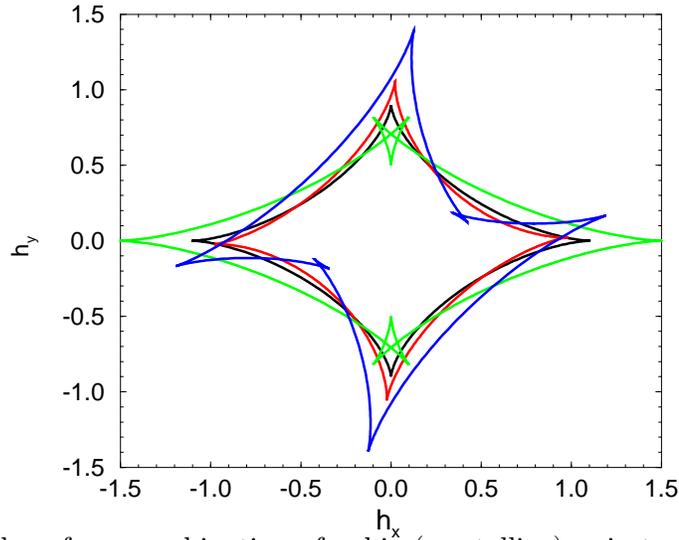,width=8cm,angle=270} \\[0.2cm]
\caption{Stability boundary for a combination of cubic (crystalline) 
anisotropy with uniaxial (shape) anisotropy. Moving from the curve with
largest width in $y-$direction to the one with smallest width in $y$
direction, the curves correspond to the parameters 1.) $c_i=0.5$,
$\gamma_i=0.5$, 2.) $c_i=0.1$, $\gamma_i=0.5$, 3.) $c_i=0.1$,
$\gamma_i=0$, and 4.) $c_i=0.5$, $\gamma_i=0.0$,
respectively.}\label{fig.bound}   
\end{figure}
\end{center}
\end{minipage}
\vspace*{0.1cm}

As is evident from eqs.(\ref{astcix}) and (\ref{astciy}), the average
astroid for fixed parameter $\theta$ is again the ideal astroid. The
standard deviations of the switching fields are given by
\begin{equation} \label{sigmach}
\sigma_{Hx,Hy}=\frac{H_c}{8}\frac{1}{\sqrt{\nu}}\sqrt{17\pm 15\cos 2\theta}\,,
\end{equation}
with the plus sign pertaining to $\sigma_{Hx}$, and the minus sign to
$\sigma_{Hy}$. Note that the total uncertainty of the switching field
$\sqrt{\sigma_{Hx}^2+\sigma_{Hy}^2}$ is independent of the parameter
$\theta$ for both cubic and uniaxial crystalline anisotropy.

\section{Numerical Simulations}\label{chap.num}
I performed micromagnetic simulations for many different cell types, varying
cell size, aspect ratio, thickness, and material properties. A commercially
available  simulation package  was used
that allows to mimic a poly--crystalline structure with an adjustable
average size of the crystallites. Each crystallite is
then assigned a random preferred direction. Typically, 10 to 100 different
disorder realizations were used for each type of cell simulated, and average
value and standard deviation of the switching fields in easy direction
(actually under a small angle of 4 degrees with respect to the easy axis in
order to avoid the numerical problems related to catastrophic switching) were
calculated. Fig.\ref{fig.relerr} shows a cumulative plot for all samples
with uniaxial or cubic
crystalline anisotropy of the standard deviation of the switching field (in
units of $K_1$) as
function of the parameter $\nu$.\footnote{Note that a negative anisotropy constant
corresponds to a preferred axis rotated by 90 degrees. As the crystallite
axes 
are uniformly  distributed over the full $\pi$ interval, $P(K,\gamma)$
should not depend on the sign of $\Kcr$. This was checked numerically by
using some samples with negative $\Kcr$.} For the calculation of $\nu$, the
numerically determined average value of $K_1$ was used, related to the
switching field by $H_c=2K_1/M_s$.
For both types of crystalline anisotropy the decay of the fluctuations like
$1/\sqrt{\nu}$ is well 
observed over almost five orders of magnitude.  For very large $\nu$ the
fluctuations seem to decay slightly slower, but they might be limited by the
finite field resolution, as well as the intrinsic fluctuations of the
simulation program. 
Ideally one would expect all curves to collapse on a single one. The
simulations show  that 
the numerical constant $\alpha$ in the definition of $\nu$ does depend
somewhat on the nominal sample properties.
\noindent
\begin{minipage}{7.0truein}
\begin{center}
\begin{figure}[h]
\epsfig{file=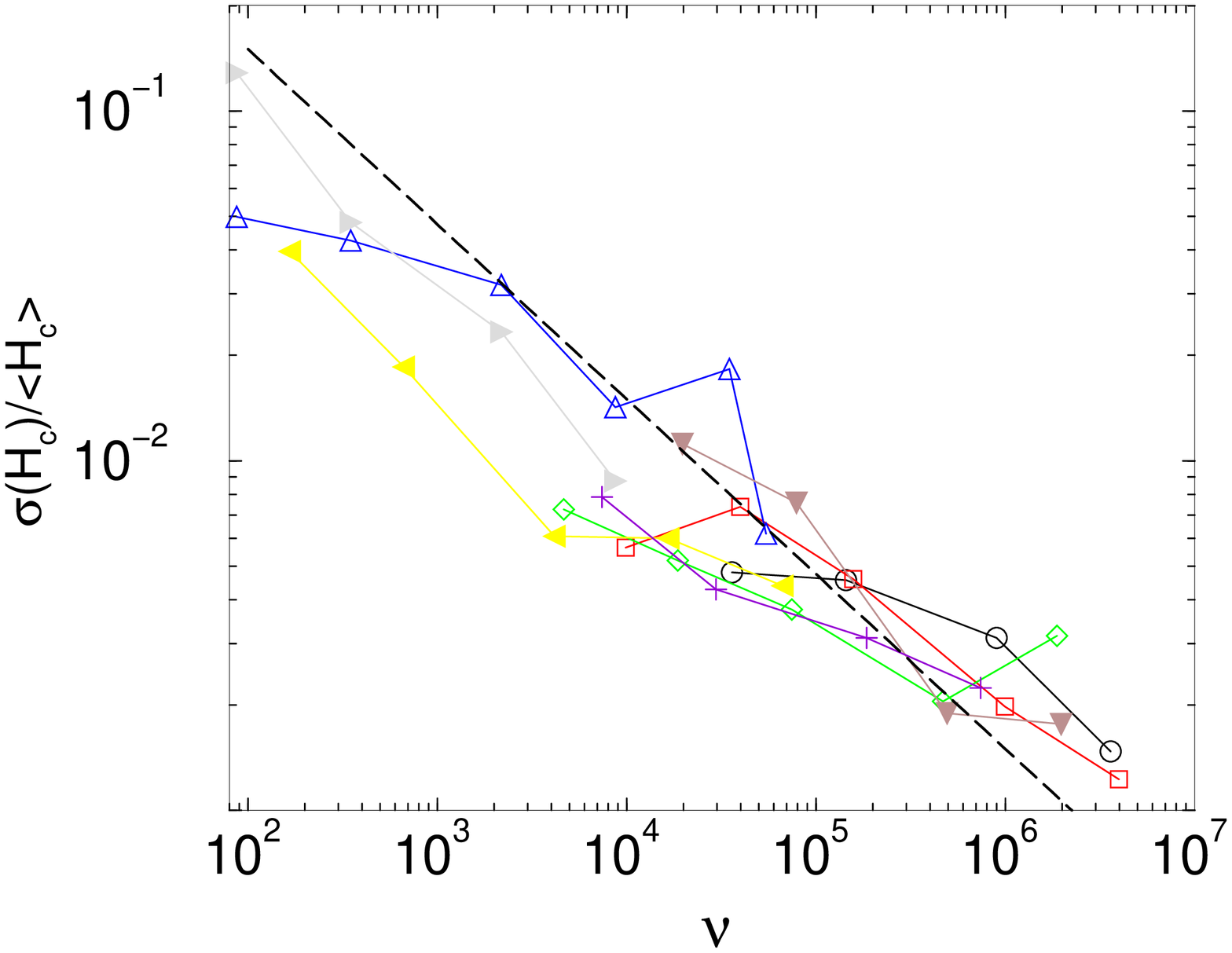,width=8cm,angle=270} \\[0.2cm]
\caption{Standard deviation of the switching field fluctuations in units of
the average switching field for various systems. Empty circles: Ellipses
$200\times400\times4\nm^3$ with 
$\Kcr=1.0 \cdot 10^3 \erg/\cm^3$; empty squares: ellipses
$300\times600\times4\nm^3$, 
$\Kcr=1.0 \cdot 10^3 
\erg/\cm^3$; empty diamonds: ellipses $300\times450\times4\nm^3$, $\Kcr=1.0
\cdot 10^3 
\erg/\cm^3$; 
empty triangles up: ellipses $200\times400\times4\nm^3$, $\Kcr=-15 \cdot
10^3 \erg/\cm^3$; 
full triangles left: 
rectangles $200\times600\times4\nm^3$, uniaxial $\Kcr=-15 \cdot 10^3
\erg/\cm^3$; full 
triangles down: $200\times400\times4\nm^3$, uniaxial $\Kcr=1.0 \cdot 10^3
\erg/\cm^3$; full 
triangles right: ellipses $200\times400\times4\nm^3$, uniaxial $\Kcr=-15
\cdot 10^3 
\erg/\cm^3$; pluses: ellipses $300\times600\times4\nm^3$, uniaxial $\Kcr=1.0
\cdot 10^3 
\erg/\cm^3$. The lines are guides to the eye for data of the same nominal
system, for which the crystallite size was varied, typically between 5 or
10nm up to 100nm. The dashed straight line indicates the $1/\sqrt{\nu}$
behavior.}\label{fig.relerr}  
\end{figure}
\end{center}
\end{minipage}
\vspace*{0.1cm}
 
\section{Conclusions}\label{chap.concl}
I have presented a study of the influence of crystalline disorder on the
switching behavior of small magnetic cells. Within a Stoner--Wohlfarth model
with random anisotropy contributions I have derived the joint--probability
distribution of the overall anisotropy strength and direction. The form of
the distribution implies a dependence of the switching field fluctuations on
a single parameter $\nu$, see eq.(\ref{nu}), in the form of a $1/\sqrt{\nu}$
behavior. Also, the width of the transition in rotational remanence
experiments should scale as $1/\sqrt{\nu}$, and a broadening of the
transition due to crystalline disorder should lead to a rather remarkable
line shape. Micromagnetic simulations confirmed the scaling with $\nu$  both
for uniaxial  
and cubic crystalline anisotropy. 

{\em Acknowledgements:} It is my pleasure to thank David Abraham, Snorri
Ingvarsson, Roger Koch, Yu Lu, John
Slonczezwski, Solomon Woods and Philip Trouilloud for stimulating
discussions. This work was done within the Magnetic RAM Developement
Alliance (MDA) between Infineon Technologies AG and the IBM Corporation.

%\end{multicols}

\end{document}